\documentclass[twoside,twocolumn,tightenlines, showpacs,aps,prb]{revtex4}
\usepackage{graphicx}
\usepackage{mathrsfs}
\usepackage{cancel}
\newcommand{\be}{\begin{equation}}
\newcommand{\ee}{\end{equation}}

\begin{document}

\title{Entanglement Distribution Statistic in Andreev Billiards}

\author{J.G.G.S. Ramos$^1$, A. F. Macedo-Junior$^2$, A.L.R. Barbosa$^2$ }

\affiliation{$^1$Departamento de F\'{\i}sica, Universidade Federal da Para\'iba, 58297-000 Jo\~ao Pessoa, Para\'iba, Brazil\\
$^2$ Departamento de F\'{\i}sica,Universidade Federal Rural de Pernambuco, 52171-900 Recife, Pernambuco, Brazil}

\date{\today}

\begin{abstract}
We investigate statistical aspects of the entanglement production for open chaotic mesoscopic billiards in contact with superconducting parts, 
known as Andreev billiards. The complete distributions of concurrence and entanglement of formation are obtained by using the Altland-Zirnbauer symmetry classes of circular ensembles of scattering matrices, which complements previous studies in chaotic universal billiards 
belonging to other classes of random matrix theory. Our results show a unique and very peculiar behavior: 
the realization of entanglement in a Andreev billiard always results in non-separable state, regardless of the time reversal symmetry. 
The analytical calculations are supported by a numerical Monte Carlo simulation.

\end{abstract}
\maketitle

\section{Introduction}

Ballistic electronic transport through chaotic mesoscopic billiards has attracted much attention in the last two decades both 
in theoretical and experimental investigations [\onlinecite{beennaker_rev, Mello_book, Gustavsson_Rev,Heinzel}]. 
Quantum interference effects inside these devices, due to multiple wave scattering at their irregular boundaries, give rise to peculiar effects
such as universal conductance fluctuations and weak localization [\onlinecite{Ramos,Dietz}].
More recently, an open chaotic ballistic billiard (CBB) has been suggested as an orbital entangler for two nointeracting electrons [\onlinecite{BeenakkerMarcus}].

The transport properties of a CBB can be described by its scattering matrix [\onlinecite{beennaker_rev, Mello_book}]. The presence of chaotic
dynamics drives the system to a universal regime for which the statistical properties of the observables can emerge from the
random matrix theory (RMT) [\onlinecite{Mehta}]. If the coupling to the outside is made 
by ideal point contacts, the scattering matrix is modeled by a random unitary matrix dropped by one of the circular ensembles 
[\onlinecite{beennaker_rev, Mello_book}]. The RMT approach is insensitive to irrelevant microscopic details of the system, but is 
strongly affected by the intrinsic symmetries of the corresponding hamiltonian, such as time-reversal (TRS), spin-rotation (SRS), 
particle-hole (PHS) and sublattes/chiral (SLS) symmetries. The presence or absence of these symmetries give rise to ten universality 
classes [\onlinecite{Jacquod}], which can be divided in three categories: (i) Wigner-Dyson (3 classes), (ii) Chiral (3 classes) and (iii)
Atland-Zirnbauer (4 classes). These categories have been used  to describe three kinds of CBB, known as (i) Schr\"odinger (SB), (ii) Dirac (DB) and
(iii) Andreev billiards (AB), respectively. 

The conductance of a CBB connected to electron reservoirs by ideal leads is one of the most important examples of transport observable 
that has been studied in all universality classes of RMT. Its average, variance and probability distributions were obtained for SB 
in Ref. [\onlinecite {Wigner_Dyson_CE, Jalabert_Pichard_Beenakker}], DB in Refs. [\onlinecite{Barros,Chiral_CE1,Nascimento }] and AB in Ref. [\onlinecite{Dahlhaus}].  

Current works have propounded several technological applications for CBB beyond electronic transport. 
Beenakker {\it et al.} [\onlinecite{BeenakkerMarcus}] suggested that it is possible to use a SB as an orbital entangler
of two non-interacting electrons. Entanglement is one the most fundamental effects of quantum mechanics with no classical
correspondence [\onlinecite{Sakurai}]. If two or more particles are entangled, the scathing meaning is their non-local 
correlations  which cannot be acquired by concepts of classical mechanics [\onlinecite{Alber}]. 
Using a model of two-channel leads, it was shown in Ref.[\onlinecite{BeenakkerMarcus}] that the average and variance of entanglement measures (concurrence and entanglement formation) are almost independent on whether TRS is present or not. 
This means that the quantum interference corrections (weak-localization) of two entanglement measures are approximately null in contrast with
another physical observables of electronic transport such as conductance. Nevertheless, in view of large fluctuations in concurrence characteristic of the extreme quantum limit, these first two moments do not reveal all relevant information about the entanglement statistics. The complete information is embedded in the full probability distribution, which was studied in Ref.[\onlinecite{Gopar}].  More recently, effects of tunneling barriers [\onlinecite{Souza,Vivo,Jarosz}] on statistics 
of concurrence and squared norm [\onlinecite{Vivo,Novaes}] for a SB were considered.

Motivated by Refs.[\onlinecite{BeenakkerMarcus,Jacquod,Gopar}], the authors of Ref.[\onlinecite{ Silva}] used the Chiral ensembles 
to study the entanglement of two electrons leaving a DB through opposing leads under the presence or not of TRS. They obtained exact
expressions for the concurrence distribution probability and showed that the average of entanglement is strongly affect by presence or not of 
TRS, in contrast of the SB behaviour. Despite the results for the Wigner-Dyson and Chiral ensembles, a study of this kind is still missing
for the Altland-Zirnbauer ensembles. In this work, we consider the concurrence statistics for the entanglement of two electrons leaving a AB (a CBB in contact with superconductor) [\onlinecite{Altland,Perez}]
through opposing leads. We calculate the exact concurrence probability distributions with and without TRS and show that they are quite 
distinct compared with both the SB [\onlinecite{Gopar}] and the DB [\onlinecite{Silva}] results. 
Furthermore, we show a TRS independence for the average of entanglemente exactly as in SB and in contrast of DB. 
Our analytical results are supported by numerical Monte Carlo simulations. 

\section{Entanglement Model}

\begin{figure}[!]
\includegraphics[width=7cm,height=5cm]{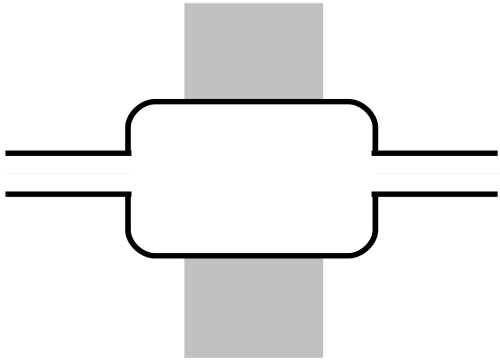} \quad \quad \quad
\caption{Open Andreev billiard: a chaotic normal metal billiard in contact with superconductor parts (grey regions) and with two pair of leads 
each one with one open channel. }\label{figura}
\end{figure}

The entanglement model for a pair of noninteracting electrons leaving of an orbital entangler was suggested by Refs.[\onlinecite{Emary,BeenakkerMarcus}]. The system consists of a CBB connected to two pair of ideal leads, each one with a single open channel, in the absence of electron-electron 
interactions at zero temperature, see Fig. \ref{figura}. An electron transmitted (absorbed) by a source (drain) through a pair of leads 
can interact directly between a member of the pair by means of a direct quantum tunneling, as the pair is settled very near.
Accordingly, before (after) the interaction within the orbital entangler, the electron is a mixed state for both the incident and the scattered
waves. The entanglement between the particles inside the CBB happens exclusively through interference effects.

The concurrence of two electrons after they leave the device can be written in terms of the eigenvalues $\tau_1$ and $\tau_2$ of the
product between the transmission block of scattering matrix and its transpose conjugate, $tt^{\dag}$, as
\begin{eqnarray}
\mathcal{C}=2\frac{\sqrt{\tau_1(1-\tau_1)\tau_2(1-\tau_2)}}{\tau_1+\tau_2-2\tau_1\tau_2}.\label{C}
\end{eqnarray}
The Eq.(\ref{C}) for the concurrence indicates that the state of the electrons as they leave the CBB is separable (or no entangled),
$\mathcal{C}=0$, if $\tau_1=1$ and/or $\tau_2=1$, while they are found maximally entangled $\mathcal{C}=1$ (Bell state) if $\tau_1=\tau_2$. 
For intermediate values of $\mathcal{C}$ between $0$ and $1$, the states are known as non-separable or partly entangled [\onlinecite{Souza}].

The entanglement of formation and the concurrence are related through [\onlinecite{Wootters}]
\begin{eqnarray}
\mathcal{E}(\mathcal{C})=h\left(\frac{1+\sqrt{1-\mathcal{C}^2}}{2}\right), \label{E}
\end{eqnarray}
where
\begin{eqnarray}
h(x)=-x \log_2(x) - (1-x)\log_2(1-x).
\end{eqnarray}
Using the Eq.(\ref{C}) and the apropriate joint transmission eigenvalues $P(\tau_1,\tau_2)$, it was showed in Ref.[\onlinecite{Gopar}] the following
expressions for the concurrence probability distribution for a SB
\begin{eqnarray} \label{pcsb1}
\mathcal{P}(\mathcal{C})&=&\frac{2}{(1+\mathcal{C})^2},
\end{eqnarray}
with TRS, and
\begin{eqnarray}
\mathcal{P}(\mathcal{C})&=&\frac{2\mathcal{C}}{(1-\mathcal{C}^2)^3}\left[3\left(2+3\mathcal{C}^2\right)\text{arctanh}\sqrt{1-\mathcal{C}^2}\right.\nonumber\\
&- &\left.\left(11+4\mathcal{C}^2\right)\sqrt{1-\mathcal{C}^2}\right]. \label{pcsb2}
\end{eqnarray}
 without TRS.
The equivalent results for a DB were obtained in Ref.[\onlinecite{ Silva}].  In the presence of TRS, the probability distribution is given by
\begin{eqnarray} \label{pcdb1}
\mathcal{P}(\mathcal{C})&=&\frac{1}{2}\frac{1}{1-\mathcal{C}^2}\left\{\frac{\sqrt{2}}{2}\sqrt{1+\frac{\sqrt{1-\mathcal{C}^2}}{1-\mathcal{C}^2}}\right. \\
&\times&\left.\textrm{arccoth}\left[ {\sqrt{1+\frac{\mathcal{C}^2}{2\left(1-\mathcal{C}^2\right)\left(1+\frac{\sqrt{1-\mathcal{C}^2}}{1-\mathcal{C}^2}\right)}}}\right]-1 \right\}\nonumber
\end{eqnarray}
and, in absence of TRS, is
\begin{eqnarray}
\mathcal{P}(\mathcal{C})&=&\frac{\sqrt{1-\mathcal{C}^2}}{(1+\mathcal{C})(1-\mathcal{C}^2)}. \label{pcdb2}
\end{eqnarray}
The Eqs.(\ref{pcsb1}), (\ref{pcsb2}), (\ref{pcdb1}) and (\ref{pcdb2}) are plotted in the Fig. \ref{figura1} to direct comparison
with the concurrence probability distribution for a AB, which will be calculated in the next section. 

\section{Probability Distributions of Andreev Billiard} 

In order to obtain statistical information of concurrence for a chaotic Andreev billiard, we use the joint transmission eigenvalues distribution 
obtained in the Ref.[\onlinecite{Dahlhaus}]. Whenever there are two transmission eigenvalues, the distribution can be written as
\begin{eqnarray}
P_{\beta,\gamma}(\tau_1,\tau_2)&=&c_{\beta,\gamma} |\tau_1-\tau_2|^{\beta} \left(\tau_1\tau_2\right)^{\beta/2-1} \times \nonumber \\
&\times& \left[\left(1-\tau_1\right)\left(1-\tau_2\right)\right]^{\gamma/2}. \label{pphi}
\end{eqnarray}
If the entangler is in presence of both SRS and TRS, the labels assume the values $\beta=2$ and $\gamma=1$, whereas, in the presence of SRS and absence of TRS,
they assume the values $\beta=4$ and $\gamma=2$. The normalization constant assumes the values $c_{2,1}=525/32$ and $c_{4,2}=1400$.

\begin{figure}[!]
\includegraphics[width=7cm,height=11cm]{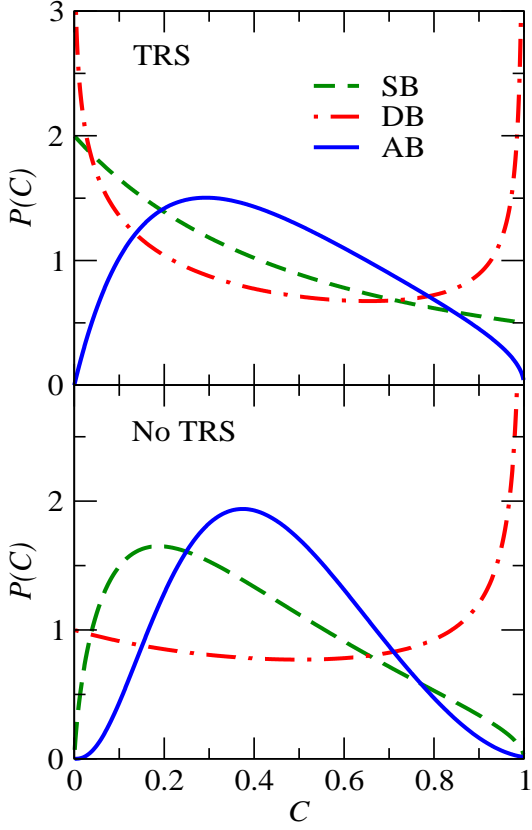} \quad \quad \quad
\caption{ The concurrence probability distributions $\mathcal{P}(\mathcal{C})$ with  and without TRS for AB, Eqs. (\ref{pc1} ) and (\ref{pc2}), 
were plotted together with the corresponding distributions for  SB [\onlinecite{Gopar}], Eqs. (\ref{pcsb1}) and (\ref{pcsb2}), and DB  [\onlinecite{Silva}], Eqs. (\ref{pcdb1}) and (\ref{pcdb2}).}\label{figura1}
\end{figure}

From Eqs.(\ref{C}) and (\ref{pphi}), we can calculate the concurrence distribution $\mathcal{P}(\mathcal{C})$ using the following definition
\begin{eqnarray}
\mathcal{P}_{\beta,\gamma}(\mathcal{C})&=&\left\langle\delta\left[\mathcal{C}-2\frac{\sqrt{\tau_1(1-\tau_1)\tau_2(1-\tau_2)}}{\tau_1+\tau_2-2\tau_1\tau_2}\right] \right\rangle,\label{pcdelta}
\end{eqnarray}
where $\left\langle\dots\right\rangle$ is the ensemble average performed with the joint eigenvalues distribution (\ref{pphi}).
By using the transformation of variables $z_i=\tau_i/(1-\tau_i)$ we can rewrite the Eq.(\ref{pcdelta}) as
\begin{equation}
\mathcal{P}_{\beta,\gamma}(\mathcal{C})=\int\!\!\!\int\delta\left[\mathcal{C}-2\frac{\sqrt{z_1z_2}}{z_1+z_2}\right]\mathcal{W}_{\beta,\gamma}(z_1,z_2)dz_1 dz_2, \label{pcphi}\qquad 
\end{equation}
where
\begin{equation}
 \mathcal{W}_{\beta,\gamma}(z_1,z_2) = \frac{\mathcal{P}_{\beta,\gamma}\left(\frac{z_1}{1+z_1},\frac{z_2}{1+z_2}\right)}{(1+z_1)^2(1+z_2)^2}.
\end{equation}

\begin{figure}[!]
\includegraphics[width=7cm,height=11cm]{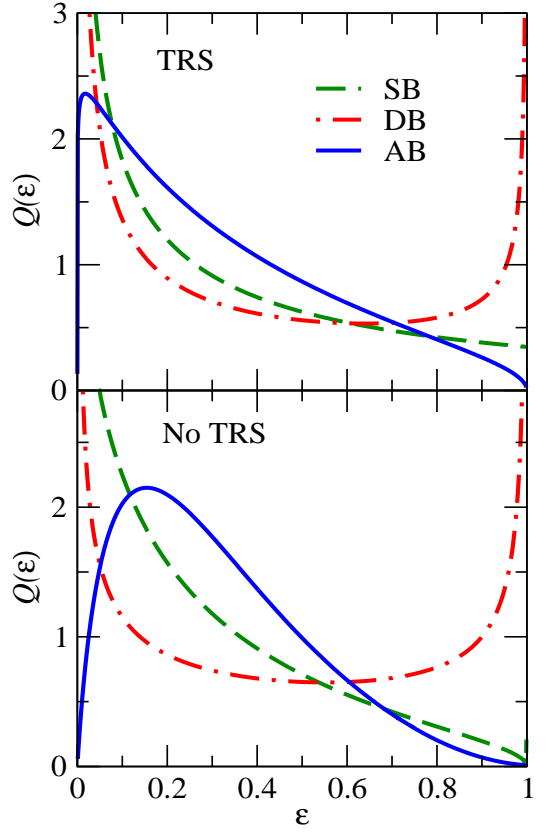} 
\caption{ The entanglement of formation probability distributions $\mathcal{Q}_\beta(\mathcal{E})$ 
from Eqs.(\ref{Q}), (\ref{pc1}) and (\ref{pc2}) were plotted for systems with and without TRS together with  the corresponding distributions for SB and DB.}\label{figura2}
\end{figure}
To perform the double integral (\ref{pcphi}), it is convenient to use the following expansion for the delta function [\onlinecite{Vivo}]
\begin{eqnarray}
\delta\left[\mathcal{C}-2\frac{\sqrt{z_1z_2}}{z_1+z_2}\right]&=& \mathcal{F}_+(\mathcal{C}) z_2\delta\left[z_1-f_+(\mathcal{C}) z_2\right]\nonumber\\&+&\mathcal{F}_-(\mathcal{C}) z_2\delta\left[z_1-f_-(\mathcal{C}) z_2\right],\quad
\end{eqnarray}
where
\begin{eqnarray}
\mathcal{F}_\pm(\mathcal{C})&=&\frac{\left(f_\pm(\mathcal{C})+1\right)^2\sqrt{f_\pm(\mathcal{C})}}{\pm f_\pm(\mathcal{C})\mp 1}
\nonumber\\
f_\pm(\mathcal{C})&=&\frac{2-\mathcal{C}^2\pm 2 \sqrt{1-\mathcal{C}^2}}{\mathcal{C}^2}.
\nonumber
\end{eqnarray}

\begin{table*}[htb]
 \centering
\begin{tabular}{ ccccccccc}
\hline
    \hline
    Ensembles \quad \quad \quad& Chaotic Billiard \quad \quad \quad &TRS \quad \quad\quad  & $\left\langle\mathcal{C}\right\rangle$\quad \quad\quad   &  $\left\langle\mathcal{C}\right\rangle_{wl}$ \quad \quad\quad  &$\textrm{var}\left[\mathcal{C}\right]$\quad \quad \quad &$\left\langle\mathcal{E}\right\rangle$ \quad \quad \quad&$\left\langle\mathcal{E}\right\rangle_{wl}$ \quad \quad\quad &$\textrm{var}\left[\mathcal{E}\right]$\\
    \hline
    \hline
 Wigner-Dyson & Schr\"odinger & YES & 0.386   & 0.0 &0.078 & 0.285 & 0.0 & 0.078\\
                         & & NO & 0.387  & & 0.056 & 0.273 &  & 0.056\\
                         \hline
 Chiral  &Dirac           & YES & 0.467   & -0.1 & 0.111 & 0.382 &-0.1 & 0.121\\
                       & & NO & 0.570   &  & 0.103 & 0.485 & & 0.122\\
                        \hline
 Altland-Zimbauer & Andreev & YES & 0.438  & 0.0 & 0.055  & 0.321 & 0.0 &0.060\\
                             &  & NO & 0.438   &  & 0.034  &0.311 & &0.038\\
\hline
\hline
\end{tabular}
\caption{ Average, variance and weak-localization correction of the concurrence and entanglement of formation for SB, DB and AB. }\label{tabela}
\end{table*}

After a some algebra, we obtain the following expressions to concurrence probability distribution if TRS is preserved 
\begin{eqnarray} \label{pc1}
\mathcal{P}_{2,1}(\mathcal{C})&=&15\frac{\mathcal{C} (1-\mathcal{C})}{\sqrt{1-\mathcal{C}^2}(1+\mathcal{C})^3}
\end{eqnarray}
and, if TRS is broken (Non-TRS),
\begin{eqnarray}
\mathcal{P}_{4,2}(\mathcal{C})&=&\frac{5}{8}\frac{\mathcal{C}^3}{(1-\mathcal{C}^2)^6}\left[2\;\text{arctanh}\sqrt{1-\mathcal{C}^2}\right.\nonumber\\
&\times &\left(560+5880 \mathcal{C}^2+7350 \mathcal{C}^4+ 1225 \mathcal{C}^6\right)\nonumber\\
&-&\left(2904+16264 \mathcal{C}^2+10350 \mathcal{C}^4+512 \mathcal{C}^6\right)\nonumber\\
&\times &\left.\sqrt{1-\mathcal{C}^2}\right]. \label{pc2}
\end{eqnarray}

In the Fig.(\ref{figura1}) the concurrence distributions for a  AB, Eqs.(\ref{pc1}) and (\ref{pc2}), are plotted together with 
the same distributions for SB and DB, Eqs. (\ref{pcsb1}), (\ref{pcsb2}), (\ref{pcdb1}) and (\ref{pcdb2}). The probability distributions 
look quite different. The main characteristic of the concurrence distribution for a AB is the peculiar null probability to find two electrons
leaving of cavity in the maximally entangled states or Bell states ($\mathcal{C}=1$) as well as in minimum entangled states
($\mathcal{C}=0$), independently of the presence of TRS symmetry. Hence, for all realization, we find the pair of electrons with some degree of 
concurrence after leaving the AB. Differently what happens with the SB and the DB, whose distributions are strong affect by presence or not of TRS,
see Fig. \ref{figura1}.

The average of concurrence can be obtained from Eqs.(\ref{pc1}) and (\ref{pc2}), yielding
\begin{eqnarray}
\left\langle\mathcal{C}\right\rangle\approx
\left\{ 
\begin{array}{cc}
0.4380 & \text{TRS} \\
0.4382 & \xcancel{\text{TRS}}
\end{array}\right. 
\label{mean}
\end{eqnarray}
In Table \ref{tabela} we summarize the average of concurrence for the three kind of billiards. From Eq.(\ref{mean}), the weak localization
(quantum interference correction) of concurrence can be calculated,   yielding
\begin{eqnarray}
\left\langle\mathcal{C}\right\rangle_{wl}=\left\langle\mathcal{C}\right\rangle_{\text{TRS} }-\left\langle\mathcal{C}\right\rangle_{ \xcancel{\text{TRS}} }\approx 0.
\label{wl}
\end{eqnarray}
Precisely as in SB [\onlinecite{BeenakkerMarcus}], the weak localization produced on the AB is approximately null, despite their very
distinct concurrence distribution behavior, while it is not null for the DB. Therefore, the break of TRS does not have a significant influence
on the average of concurrence until the fourth decimal for both AB and SB, while it has a strong influence on DB, as we summarize on the Table \ref{tabela}. For the variance of concurrence, we obtain
\begin{eqnarray}
\textrm{var}\left[\mathcal{C}\right]\approx\left\{ \matrix{
0.0558 \qquad \text{TRS} \cr
0.0349 \qquad \xcancel{\text{TRS} }}\right.
\label{var}
\end{eqnarray}
As the Table \ref{tabela} indicates, the variance of concurrence of SB and AB are of the same order, while it approximately doubles for the DB.

Lastly, we obtain the  entanglement of formation distribution $\mathcal{Q}_{\beta,\gamma}$ using the Eqs.(\ref{pc1}) and (\ref{pc2}) and an appropriate change
of variables [\onlinecite{Gopar}]:
\begin{eqnarray}
\mathcal{Q}_{\beta,\gamma}(\mathcal{E})&=&\frac{\ln (2)}{\mathcal{C}(\mathcal{E})}\frac{\sqrt{1-\mathcal{C}(\mathcal{E})^2}}{\textrm{arctanh} [\sqrt{1-\mathcal{C}(\mathcal{E})^2}]}\mathcal{P}_{\beta,\gamma}(\mathcal{E}). \label{Q}
\end{eqnarray}
The entanglement of formation distributions are plotted in the Fig. \ref{figura2} for the AB together with the SB and DB distributions for a direct comparison.
The concurrence and entanglement distributions in the AB has a unique and very peculiar behavior: they can never be found in Bell state
($\mathcal{E}=1$) and in separable state ($\mathcal{E}=0$). All experiments in AB give rise to
a non-separable state regardless of the time reversal symmetry.

The average of entanglement obtained from the characteristic distribution is
\begin{eqnarray}
\left\langle\mathcal{E}\right\rangle\approx\left\{ \matrix{
0.321 \qquad \text{TRS}  \cr
0.311 \qquad  \xcancel{\text{TRS}}  }\right.
\label{meane}
\end{eqnarray}
revealing the weak localization (quantum interference correction) term of the entanglement
\begin{eqnarray}
\left\langle\mathcal{E}\right\rangle_{wl}=\left\langle\mathcal{E}\right\rangle_{\text{TRS} }-\left\langle\mathcal{E}\right\rangle_{ \xcancel{\text{TRS}} }\approx 0.
\label{wle}
\end{eqnarray}
For the variance, we obtained
\begin{eqnarray}
\textrm{var}\left[\mathcal{E}\right]\approx\left\{ \matrix{
0.060 \qquad \text{TRS}  \cr
0.038 \qquad  \xcancel{\text{TRS}}  }\right.
\label{vare}
\end{eqnarray}
For comparison, all these numbers are listed in the Table \ref{tabela} for the three kind of billiards.

\section {Monte Carlo Simulation}

We can numerically calculate statistical properties of concurrence and entanglement by using a Monte Carlo approach [\onlinecite{Ashcrof}]. 
The starting point is the representation of the joint transmission eigenvalue distribution of the Eq.(\ref{pphi}) as a Gibbs distribution of
statistical mechanics
\be
\mathcal{P}_{\beta,\gamma}(\tau_1,\tau_2) = \frac{1}{\mathcal{Z}}e^{-\beta \mathscr{H}},
\ee
with symmetry index $\beta$ playing the role of inverse of temperature and hamiltonian given by
$
\mathscr{H} = v(\tau_1,\tau_2) + u(\tau_1) +u(\tau_2)
$,
where $$v(\tau_1,\tau_2) = -\ln|\tau_1-\tau_2|$$ is a 2D-Coulombian repulsive interaction and 
$$u(\tau) = \frac{2-\beta}{2\beta}\ln\tau + \frac{\gamma}{2\beta}\ln (1-\tau)$$
is a confining potential. 

\begin{figure}[!]
\includegraphics[width=7cm,height=11cm]{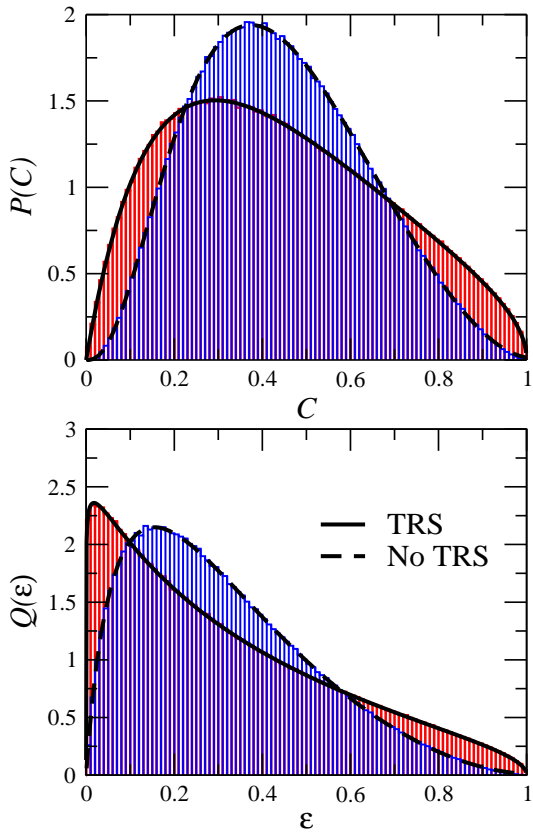} 
\caption{(Up) The concurrence distributions from Eqs. (\ref{pc1}) and (\ref{pc2}) are plotted with numerical results 
from Monte Carlos simulation. (Down) The entanglement of formation  distributions from Eq. (\ref{Q}) are plotted with numerical results from Monte Carlos simulation. In both distributions we obtain the excellent agreement between simulation and analytical results.}\label{figura3}
\end{figure}

Such Coulomb gas analogy allows for a Monte Carlo simulation. We used a Metropolis algorithm [\onlinecite{Metropolis}] to generate 
acceptable equilibrium configurations $\tau_1$ and $\tau_2$ from which we calculated the observables $\mathcal{C}$ and $\mathcal{E}$. 
In the Fig. \ref{figura3} we show the excellent agreement between simulation and analytical results for concurrence and entanglement distributions in cases with and without TRS.

\section {Discussions and Conclusions} 

In this work, we present a complete statistical study of the electronic entanglement in a chaotic Andreev Billiard.
The results are compared with pervious ones which analyzed the electronic entanglement in the chaotic Schr\"odinger and Dirac Billiards. 
Our analytical findings indicate that the concurrence and also entanglement probability distributions in the AB have a unique and very peculiar
behavior, they are non-separable states. This means that all realizations of entanglement in a AB are non-separable, regardless of the time 
reversal symmetry. The results are numerically supported by Monte Carlo simulation.

Furthermore, the presented results close the analysis of entanglement in a CBB in the framework of universal RMT. 
We explicitly compare all the universal orbital entanglers and show some of their peculiarities with analytical
results such as average, weak localization correction and variance of the concurrence and entanglement of formation. 
We believe that our work contributes to a better understanding of electronic entanglement in the framework of RMT and may give rise experimental
perspectives on the use of the AB as an orbital entangler.

This work was partially supported by CNPq, CAPES and FACEPE (Brazilian Agencies).

\end{document}